\documentclass[aps,pra,twocolumn,groupedaddress,showpacs]{revtex4-1}
\usepackage{graphicx}												
\usepackage{color}													
\usepackage{hyperref}												

\begin{document}

\title{Optogalvanic spectroscopy of the hyperfine structure of the $5p^65d$ ${}^2D_{3/2,5/2}$ and $5p^64f$ ${}^2F^o_{5/2,7/2}$ levels of La III}


\author{S. Olmschenk}
	\email{steven.olmschenk@denison.edu}
\author{P. R. Banner}
\author{J. Hankes}
\author{A. M. Nelson}

\affiliation{Denison University, 100 West College Street, Granville, Ohio 43023, USA}

\date{\today}

\begin{abstract}
We measure the hyperfine structure of the $5p^65d$ ${}^2D_{3/2}$, $5p^65d$ ${}^2D_{5/2}$, $5p^64f$ ${}^2F^o_{5/2}$, and $5p^64f$ ${}^2F^o_{7/2}$ levels in doubly-ionized lanthanum (La III; La$^{2+}$) in a hollow cathode lamp using optogalvanic spectroscopy.  Analysis of the observed spectra allows us to determine the hyperfine $A$ coefficients for these levels to be $A_{D3/2}=412(4)$ MHz, $A_{D5/2}=20(5)$ MHz, $A_{F5/2}=319(2)$ MHz, and $A_{F7/2}=155(4)$ MHz; and provide estimates for the hyperfine $B$ coefficients as $B_{D3/2}=105(29)$ MHz, $B_{D5/2}=157(40)$ MHz, $B_{F5/2}=-2(53)$ MHz, and $B_{F7/2}=171(51)$ MHz.
\end{abstract}


\maketitle

\section{Introduction}
\label{sec:intro}
The atomic structure of lanthanum ions is of interest to astrophysical measurements of stellar composition~\cite{cowley1984:cpstars, lawler2001:lifetimes_branching_osc_LaII_solar, wahlgren2002:lanthanides_stellar, jorissen2004:amo_stellar}, appraisals of atomic structure calculations for atomic clocks and variations of fundamental constants~\cite{dzuba2012:ion_clock_alpha, safronova2014:laiii, safronova2014:hci_clock_alpha, safronova2015:correlation_lanthanide_ions}, measurements of parity nonconservation~\cite{roberts2013:pnc_fr_cs}, and a proposal for laser cooling and quantum information~\cite{olmschenk2014:lanthanum_proposal}.  The hyperfine structure of singly-ionized lanthanum has been investigated with a range of techniques~\cite{furmann2010:tech_laser_spect_lanthanides, windholz2017:hyperfine_laii}, including experimental observations using grating spectroscopy~\cite{meggers1927:hyperfine_lanthanum}, interferometry~\cite{luhrs1955:hyperfine_lanthanum}, collinear ion-beam-laser spectroscopy~\cite{hohle1982:hyperfine_laii, maosheng2000:hyperfine_laii, li2001:hyperfine_laii, hongliang2002:laii}, Fourier transform spectroscopy~\cite{lawler2001:lifetimes_branching_osc_LaII_solar, guzelcimen2013:fts_lanthanum}, a laser and radiofrequency double resonance technique~\cite{schef2006:hyperfine_laii}, and laser-induced fluorescence~\cite{furmann2008:laii_odd, furmann2008:laii_even, nighat2010:hyperfine_la_lif}, as well as theoretical calculations using a classical parametric scheme~\cite{bauche1982:hyperfine_laii}, a relativistic configuration-interaction method~\cite{datta1995:hyperfine_laii}, and a semi-empirical method~\cite{furmann2008:laii_odd, furmann2008:laii_even}.  Although many of the parameters for doubly-ionized lanthanum (La III; La$^{2+}$) also have been investigated experimentally~\cite{gibbs1926:doublets_atoms, gibbs1929:doublets_laiii, badami1931:spect_CeIV, russell1932:lanthanum_spectra, sugar1965:laiii, odabasi1967:laiii, johansson1971:laiii, muller1989:electron_ionization_la_ions, li1999:lifetime_laii_laiii, biemont1999:laiii_luiii} and theoretically~\cite{lindgard1977:trans_prob_iso_seq, migdalek1984:rel_effects_rb_cs_seq, migdalek1987:cs_iso_seq, eliav1998:trans_energy_la_ions, quinet2004:lande_g_doubly_ion_lanthanides, karacoban2012:laiii, karacoban2012:laiii_trans_params, safronova2014:laiii}, the hyperfine structure of La$^{2+}$ is known for only a few energy levels.  Specifically, using grating spectroscopy~\cite{crawford1935:hyperfine_laiii, odabasi1967:laiii} and interferometry~\cite{wittke1940:laiii} the hyperfine structure of the metastable $6s$ ${}^2S_{1/2}$ and excited $6p$ ${}^2P_{1/2,3/2}$ levels of La$^{2+}$ were measured.  The hyperfine structure of the lowest levels, which may be strongly influenced by electron correlations~\cite{safronova2014:laiii}, has not been determined previously.  Here, we use Doppler-limited optogalvanic spectroscopy to measure the hyperfine structure of the lowest energy levels of La$^{2+}$.

Optogalvanic spectroscopy consists of monitoring the conductivity of (or the current through) a discharge illuminated with tunable light, where the optogalvanic effect results in a change in the electrical properties of a gas discharge when incident light is resonant with a constituent atomic or molecular transition~\cite{penning1928:optogalvanic, green1976:optogalvanic, goldsmith1981:optogalvanic, barbieri1990:optogalvanic}.  Discharges are widely used to interrogate the energy level structure of ions with a range of techniques, including optogalvanic spectroscopy, since collisional excitation results in population of high-lying energy levels of atoms and molecules in the discharge, including ionized states.  Additionally, sputtering from the discharge can produce gas-phase atoms, ions, and molecules from even refractory materials.  Optogalvanic spectroscopy is also a very sensitive technique, allowing for measurements of weak transitions and sparsely populated states~\cite{keller1980:optogalvanic_noise, cavasso2001:ca_hcl, siddiqui2013:lai_weak}.  

\begin{figure}
  \includegraphics[width=0.9\columnwidth,keepaspectratio]{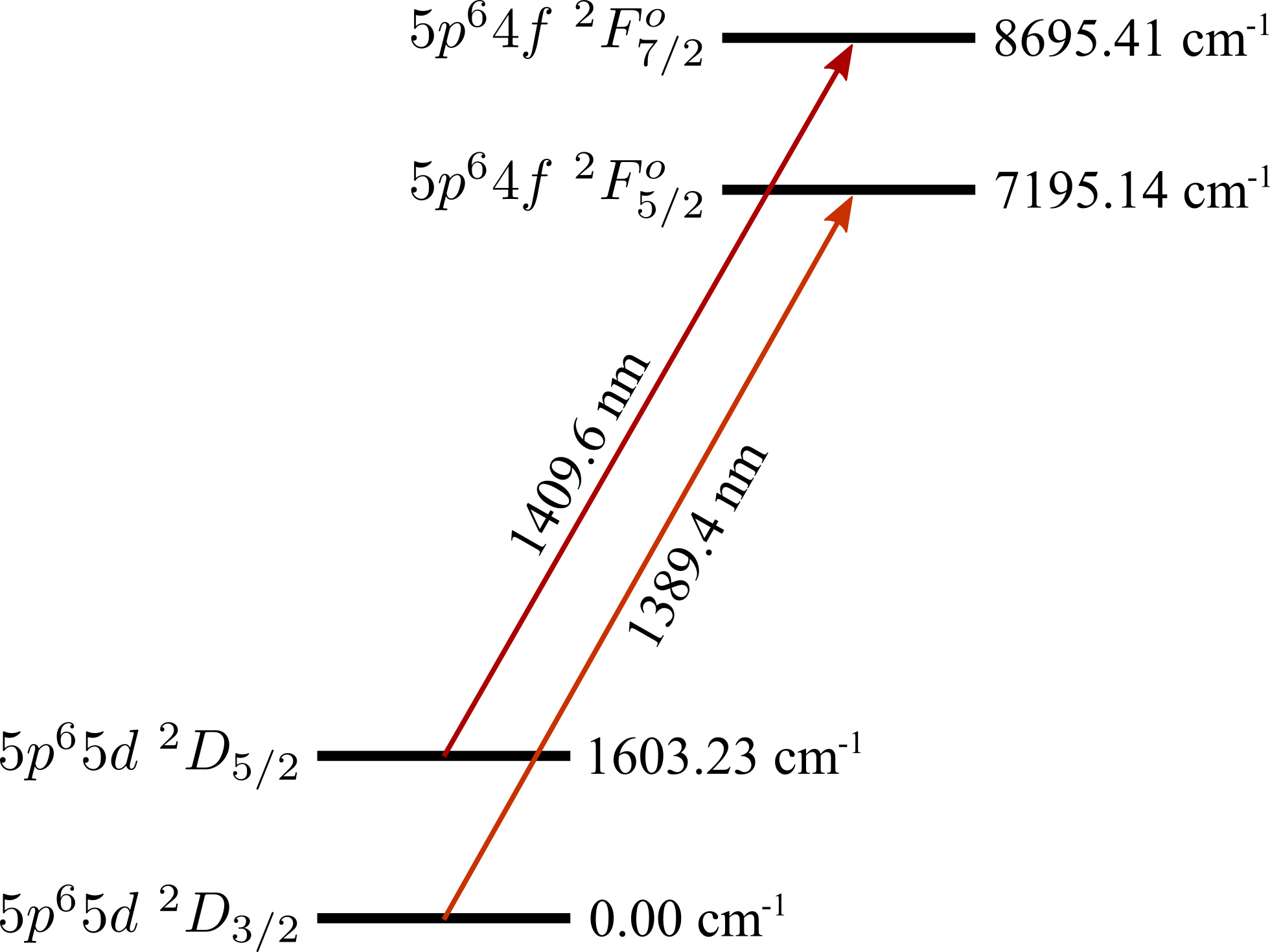}
\caption{La$^{2+}$ level diagram.  The four lowest energy levels of La$^{2+}$ are shown~\cite{NIST:ASD}, along with the transitions addressed in this experiment (wavelengths given in air).}
\label{fig:level_diagram}       
\end{figure}

In this experiment, we drive the ${}^2D_{3/2} \leftrightarrow {}^2F^o_{5/2}$ transition near 1389.4 nm (air), and the ${}^2D_{5/2} \leftrightarrow {}^2F^o_{7/2}$ transition near 1409.6 nm (air), in La$^{2+}$ (Fig.~\ref{fig:level_diagram}) and measure the resulting optogalvanic signal.  Analysis of the optogalvanic spectra allows us to determine hyperfine coefficients of the $5p^65d$ ${}^2D_{3/2}$, $5p^65d$ ${}^2D_{5/2}$, $5p^64f$ ${}^2F^o_{5/2}$, and $5p^64f$ ${}^2F^o_{7/2}$ levels.  As the ${}^{139}$La isotope has a natural abundance of 99.91\%~\cite{delaeter2003:atomic_weights}, all of our results are for ${}^{139}$La$^{2+}$, which has nuclear spin $I=7/2$.

%

\section{Experimental Setup}
\label{sec:expt_setup}
\begin{figure}
  \includegraphics[width=1.0\columnwidth,keepaspectratio]{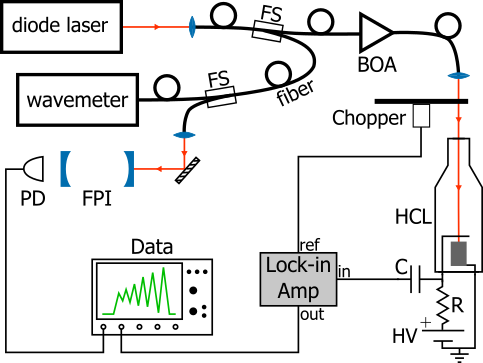}
\caption{Schematic of the experimental setup.  Light from one of two diode lasers is coupled into a port of a 50:50 single-mode fiber splitter (only one diode laser/input port shown).  A fraction of the light is directed through another fiber splitter, and subsequently to a wavemeter and Fabry-Perot interferometer.  The other portion of the light is amplified by a broadband optical amplifier, and directed through a chopper and into the bore of a hollow-cathode lamp.  The beam chopping frequency is used as the reference for a lock-in amplifier.  The optogalvanic signal is sent to the input of the lock-in amplifier.  The output of the lock-in amplifier is recorded on an oscilloscope, along with the laser transmission through the Fabry-Perot interferometer.  A voltage ramp (not shown) is used to scan the wavelength of the laser and triggers the oscilloscope.  FS is fiber splitter; BOA is broadband optical amplifier; PD is photodiode; FPI is Fabry-Perot interferometer; HCL is hollow-cathode lamp; C is 30-nF capacitor; R is 10-k$\Omega$ resistor; HV is high voltage.}
\label{fig:optogalvanic_setup}       
\end{figure}

Lanthanum ions are generated in the discharge of a hollow-cathode lamp (HCL).  The commercial HCL is single-ended, with an argon fill-gas specified at about 4 torr (Photron P827A).  A high voltage power supply (SRS PS310) drives the HCL in series with a 10-k$\Omega$ resistor, as shown schematically in Fig.~\ref{fig:optogalvanic_setup}.  The power supply is operated at 240 V, resulting in about 11.5 mA sustained through the discharge of the HCL~\cite{hcl_note}.

Laser light used to interrogate the ions is produced by two custom extended-cavity diode lasers~\cite{cook:stable_diode_laser}, one operating near 1389.4 nm and one operating near 1409.6 nm.  Light from the selected laser is coupled into a single-mode optical fiber, and subsequent fiber splitters direct a portion of this light to a wavemeter and a Fabry-Perot interferometer (FPI), as illustrated in Fig.~\ref{fig:optogalvanic_setup}.  The remainder of the light is input to a fiber-coupled broadband optical amplifier (Thorlabs BOA1036P), capable of producing more than 60 mW at each wavelength.  The output of the amplifier is directed through an optical chopper and into the bore of the HCL.

The beam is chopped (amplitude modulated) at a frequency of about 1.1 kHz, and this frequency is used as the reference for a lock-in amplifier.  A capacitor in the HCL supply circuit couples the optogalvanic-induced current modulation to the input of the lock-in amplifier.  An oscilloscope records the output of the lock-in amplifier, as well as transmission peaks through the FPI, as the laser wavelength is scanned across a La$^{2+}$ transition.  The resulting optogalvanic spectrum is averaged over either 5 or 10 scans, and subsequently stored and transferred for analysis.

\section{Data and Analysis}
\label{sec:data_analysis}
Optogalvanic spectra peak positions are determined by the hyperfine energy shifts of the investigated levels.  In terms of the hyperfine $A$ and $B$ coefficients, the energy shifts $\Delta E_{hf}$ are given by~\cite{arimondo1977:hyperfine}
\begin{equation}
	\label{eq:hfs_shift}
		\Delta E_{hf} = \frac{h}{2} A K + h B \frac{\frac{3}{2} K (K+1) - 2 I (I+1) J (J+1)}{2 I (2I-1) 2 J (2J-1)}
\end{equation}
where $I$ is the nuclear spin, $J$ is the total electron angular momentum, $F = I + J$ is the total angular momentum, and $K = F(F+1) - I(I+1) - J(J+1)$.  Thus, all of the allowed transition frequencies between the two levels are determined by a set of hyperfine coefficients for each level (with known $J$ and $I$), selection rules, and a value for the unperturbed energy level difference.  The relative intensity of each transition is calculated using a Wigner 6-j symbol~\cite{emery2006:handbook_atomic_physics_hyperfine}.

\begin{figure}
  \includegraphics[width=1.0\columnwidth,keepaspectratio]{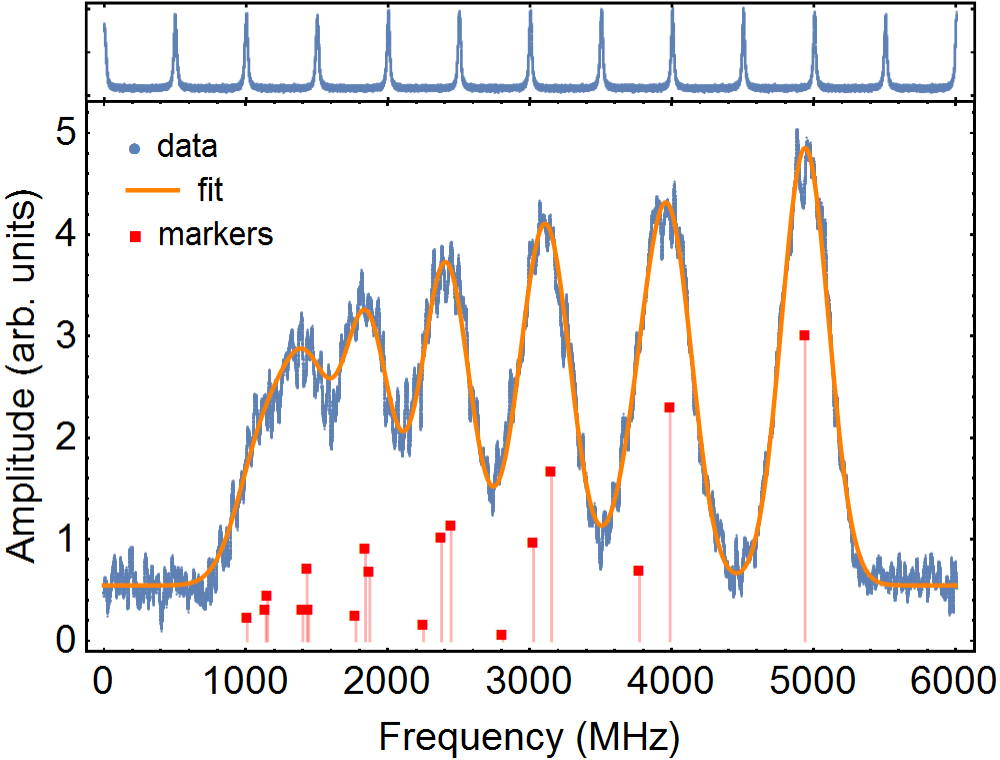}
\caption{Optogalvanic spectrum near 1409.6 nm.  \textit{Upper}: Transmission peaks through the Fabry-Perot interferometer, used to set the frequency scale (horizontal axis).  \textit{Lower}: The blue, circular points are the measured optogalvanic signal as a function of the laser detuning.  Incident laser power for this data set is about 60 mW; data shown is averaged over 5 scans.  The orange, solid line is a fit to the data, with parameters as described in the text.  The red, square, vertical markers indicate the fitted position of each hyperfine transition, with the height of the marker representing the relative intensity of the transition.  These transitions, from left (lower frequency) to right (higher frequency), in terms of $F_{lower}$,$F_{higher}$ are:  1,0; 2,1; 1,1; 3,2; 2,2; 1,2; 4,3; 3,3; 2,3; 5,4; 4,4; 3,4; 6,5; 5,5; 4,5; 6,6; 5,6; 6,7.}
\label{fig:og_14096_spectrum_v2}       
\end{figure}
%

\begin{figure}
  \includegraphics[width=1.0\columnwidth,keepaspectratio]{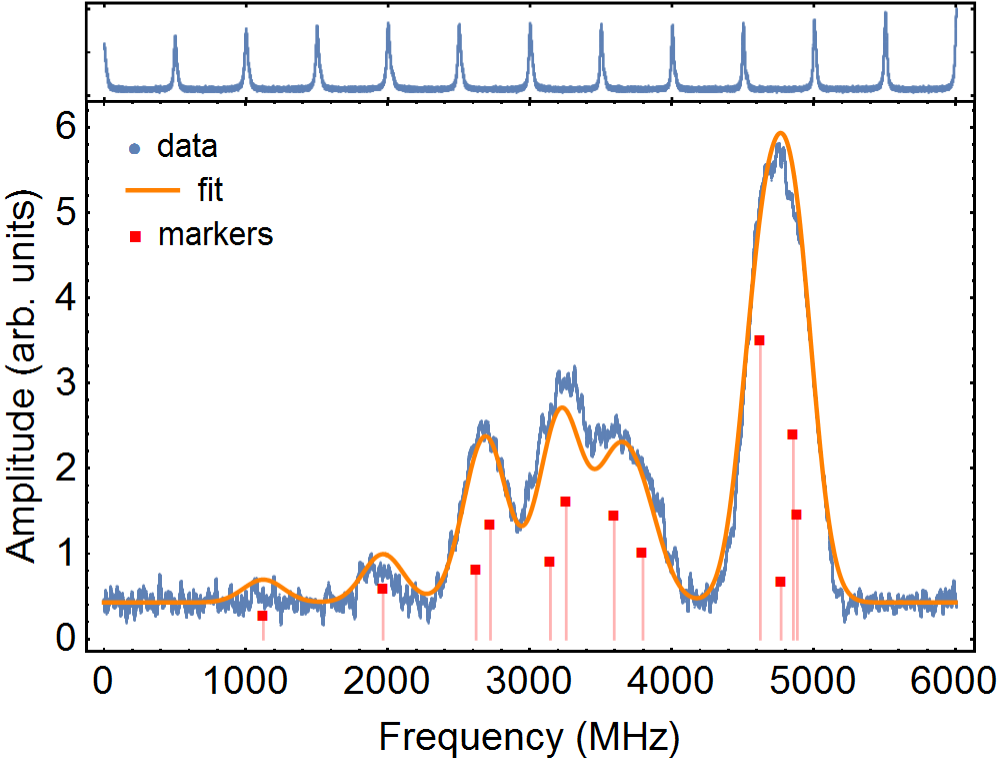}
\caption{Optogalvanic spectrum near 1389.4 nm.  \textit{Upper}: Transmission peaks through the Fabry-Perot interferometer, used to set the frequency scale (horizontal axis).  \textit{Lower}: The blue, circular points are the measured optogalvanic signal as a function of the laser detuning.  Incident laser power for this data set is about 50 mW; data shown is averaged over 10 scans.  The orange, solid line is a fit to the data, with parameters as described in the text.  The red, square, vertical markers indicate the fitted position of each hyperfine transition, with the height of the marker representing the relative intensity of the transition.  These transitions, from left (lower frequency) to right (higher frequency), in terms of $F_{lower}$,$F_{higher}$ are:  5,4; 4,3; 3,2; 5,5; 2,1; 4,4; 3,3; 2,2; 5,6; 2,3; 4,5; 3,4.}
\label{fig:og_13894_spectrum_v2}       
\end{figure}

We record spectra at incident average beam powers of 10, 20, 30, 40, 50 (twice), and 60 mW, with representative data shown in Fig.~\ref{fig:og_14096_spectrum_v2} for the 1409.6 nm transition and Fig.~\ref{fig:og_13894_spectrum_v2} for the 1389.4 nm transition.  The frequency scale is determined by the transmission peaks through the FPI.  The FPI is a custom confocal optical cavity composed of two concave mirrors with a radius of curvature of about 15 cm, mounted in an invar holder.  The free spectral range of the FPI is measured to be $500.7(2)$ MHz by using a fiber electro-optic modulator to modulate the light incident on the cavity (with frequencies up to 4.4 GHz), and optimizing the overlap of the resulting sidebands as a function of the driving frequency.  For each spectrum, we determine the position of each recorded transmission peak by a lorenzian fit, taking the distance between peak positions as the free spectral range of the FPI, and assuming a linear frequency scaling between each set of adjacent peaks.

\begin{figure}
  \includegraphics[width=1.0\columnwidth,keepaspectratio]{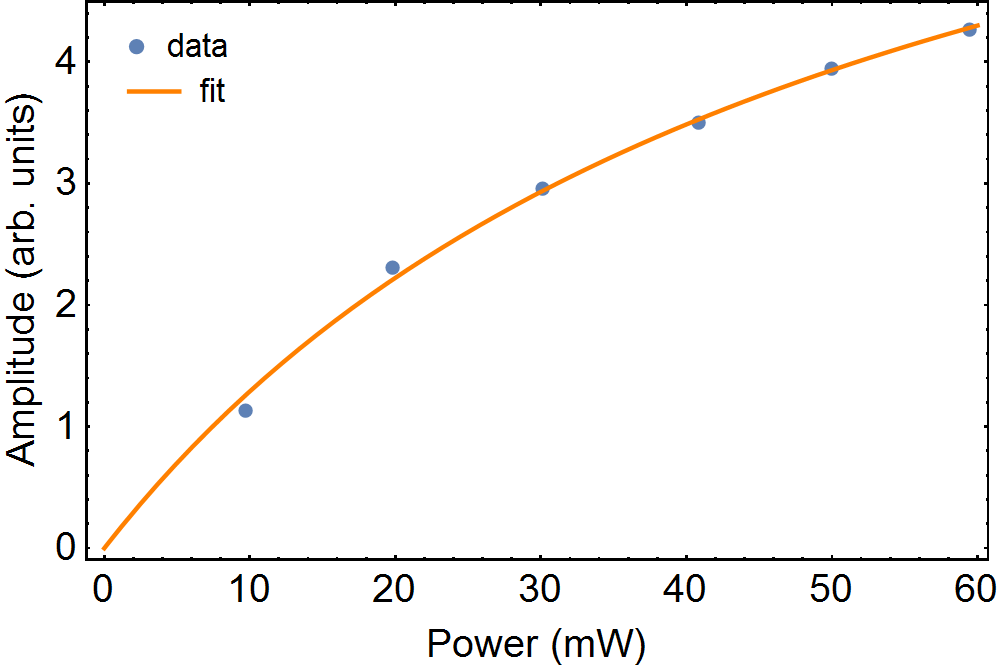}
\caption{Optical saturation of the optogalvanic signal.  A simple gaussian fit is used to determine the amplitude of a single peak in the 1409.6 nm spectrum, due to an isolated hyperfine transition, as a function of incident laser power~\cite{saturation_note}.  The solid line is a fit to $a (P/P_{sat})/\left[(P/P_{sat})+1\right]$, resulting in an estimated saturation power of $P_{sat} = 52(7)$ mW.  Given a measured beam waist of about 0.7 mm, this value for $P_{sat}$ corresponds to a peak intensity of about $7\times10^4$ W/m$^2$.}
\label{fig:saturation}       
\end{figure}

Each optogalvanic spectrum is fit to a function that uses the hyperfine $A$ and $B$ coefficients for each level as parameters to determine the transition frequencies, calculates the relative intensities, and assumes a gaussian profile of equal width for each transition~\cite{voigt_note}.  The hyperfine coefficients, gaussian width, overall laser detuning, and an overall multiplicative factor are optimized to obtain the best fit.  A background offset term is separately determined by averaging a portion of the measured signal away from the transition peaks.  As optical saturation is observed at higher incident laser power (Fig.~\ref{fig:saturation}), a saturation parameter is also included in the fitting routine, which modifies the relative intensity of the hyperfine transitions~\cite{engleman1985:optogalvanic_saturation}.

The statistical uncertainty in the hyperfine $A$ and $B$ coefficients is taken as the standard deviation of the values from all spectra for a given transition, which ranges from about 2 to 5 MHz for the $A$ coefficients, and 27 to 53 MHz for the $B$ coefficients.  The uncertainty of the fit also contributes to the overall uncertainty, and is determined by a $\chi^2$ analysis of the $A$ and $B$ coefficients, with optimized (fitted) values for all other parameters.  The fit uncertainty is less than 1 MHz for all $A$ coefficients, and less than 5 MHz for all $B$ coefficients, with the exception of the $B$ coefficients from the 1409.6 nm spectrum at 10 mW incident power, which had almost an order of magnitude larger uncertainty.  The uncertainty of the fit is also used to weight the results from each spectrum, such as for calculating the mean value of each hyperfine coefficient.

\begin{table}[b]
	\caption{Hyperfine $A$ and $B$ coefficients for the lowest levels of La$^{2+}$, with levels and energies from Ref.~\cite{NIST:ASD}.  Uncertainties in the measured hyperfine coefficients are given in parentheses after each value (in MHz), and are statistical and systematic uncertainties as described in the text, added in quadrature.}
	\label{tab:hfs}%
	\begin{ruledtabular}
		\begin{tabular}{lrrr}
			\textrm{Level}				&
			\textrm{Energy (cm$^{-1}$)}				&
			\textrm{$A$ (MHz)}		&
			\textrm{$B$ (MHz)}		\\
			\colrule
			$5p^65d$ ${}^2D_{3/2}$		& 0.00		& 412(4)	& 105(29)	\\
			$5p^65d$ ${}^2D_{5/2}$		& 1603.23	& 20(5)		& 157(40)	\\
			$5p^64f$ ${}^2F^o_{5/2}$	& 7195.14	& 319(2)	& -2(53)	\\
			$5p^64f$ ${}^2F^o_{7/2}$	& 8695.41	& 155(4)	& 171(51)	\\
		\end{tabular}
	\end{ruledtabular}
\end{table}

A possible systematic error is laser frequency drift during data acquisition.  We model this by assuming a potential laser frequency drift as large as one half-width half-max of the FPI transmission peaks over the course of a scan; since the data is averaged over 5 or 10 scans, larger drifts would be evident in the FPI data.  Reanalyzing the optogalvanic spectra with this modeled drift shifts the mean value of the $A$ coefficients by less than 1 MHz, and the $B$ coefficients by less than 3 MHz.

Another systematic error is incident laser power variation across a scan.  In a single scan, the amplitude of the FPI transmission peaks vary by as much as 48\%.  In order to evaluate the error, we fit the transmission peaks at each end of a single scan from each spectrum to determine the amplitude (power) variation, and model this effect by multiplying the optogalvanic spectrum by a linear function that varies by this amount across the spectrum.  In all cases, we impose a variation of at least 5\% to model this potential error.  Analysis of the modified spectra shows that this modeled power variation changes the mean value of the $A$ coefficients by less than 3 MHz, and the $B$ coefficients by less than 24 MHz.

The mean values for the $A$ and $B$ hyperfine coefficients for each energy level are given in Table~\ref{tab:hfs}, where the tabulated uncertainties are statistical and systematic uncertainties added in quadrature.  As is seen, the uncertainty in the $B$ coefficients is large, due to the limited resolution of the spectra.  Investigating the hyperfine coefficients using Doppler-free or Doppler-reduced techniques, such as saturated absorption spectroscopy~\cite{gough1985:hcl_sat_abs} or intermodulated optogalvanic spectroscopy~\cite{lawler1979:imogs}, will undoubtedly reduce the uncertainty in these values.

%

\section{Conclusion}
\label{sec:conclusion}
Using optogalvanic spectroscopy, we determined the hyperfine $A$ and $B$ coefficients for the four lowest energy levels of ${}^{139}$La$^{2+}$.  These measurements of the hyperfine structure of doubly-ionized lanthanum may be useful for a range of experiments in astrophysics and atomic physics, and may enable laser cooling of this ion in the future.

\begin{acknowledgments}
We thank P. Becker and E. Pewitt for early contributions to the optogalvanic setup.  P.B. and A.N. acknowledge support from the Laurie Bukovac and David Hodgson Endowed Fund at Denison University; P.B. and J.H. acknowledge support from the J. Reid \& Polly Anderson Endowed Fund at Denison University.  This material is based upon work supported by, or in part by, the U. S. Army Research Laboratory and the U. S. Army Research Office under contract/grant number W911NF-13-1-0410; Research Corporation for Science Advancement through Cottrell College Science Award 22646; and Denison University.  Specific product citations are for the purpose of clarification only, and are not an endorsement by the authors, the U. S. Army Research Laboratory, the U. S. Army Research Office, Research Corporation for Science Advancement, or Denison University.
\end{acknowledgments}

%


\begin{thebibliography}{63}
\bibitem [{hcl()}]{hcl_note}%
  \BibitemOpen
  \href@noop {} {}\bibinfo {note} {The effect of different HCL fill-gases and
  operating currents were also investigated. Comparing the La$^{2+}$
  optogalvanic signal from the argon-filled HCL to that from a neon-filled HCL,
  with similar optical alignment and operating current, indicates that the
  argon-filled HCL produces a signal more than 10 times the signal from the
  neon-filled HCL. However, the noise in the argon-filled HCL is also greater
  than the noise in the neon-filled HCL, resulting in a net signal-to-noise
  ratio (SNR) improvement of about 3. In both lamps, the optogalvanic signal
  increases with operating current, though at the cost of reduced lamp
  lifetime.}\BibitemShut {Stop}%
\bibitem [{sat()}]{saturation_note}%
  \BibitemOpen
  \href@noop {} {}\bibinfo {note} {While all spectra noted in the text are used
  to determine the hyperfine coefficients, one 50-mW spectrum from the
  1409.6-nm transition is not included in Fig.~\ref{fig:saturation}, as the
  beam alignment and HCL parameters were slightly different than the other data
  sets plotted here.}\BibitemShut {Stop}%
\end{thebibliography}
\end{document}